\documentclass[a4paper,11pt]{article}
\usepackage{jinstpub}
\usepackage{longtable}

\newcommand{\eos}{\textsc{Eos}}
\newcommand{\finemet}{\textsc{FINEMET}}
\newcommand{\deltat}{$\Delta\rm{t}$}
\newcommand{\gain}{$1 \times 10^{7}$}

\usepackage[sorting=none, style=nature]{biblatex} 
\bibliography{bibliography} 

\title{Characterization of the Hamamatsu 8-inch R14688-100 PMT}

\author[1,2]{Tanner Kaptanoglu}
\author[1]{Ashley Rincon}
\author[4]{Mackenzie Duce}
\author[1]{Sawyer Kaplan}
\author[1]{Joseph Koplowitz}
\author[3]{Skipper Lynch}
\author[1]{Hong Joo Ryoo}
\author[1,2]{Gabriel Orebi Gann}

\affiliation[1]{University of California, Berkeley, 366 Physics North, Berkeley, CA 94720-7300}

\affiliation[2]{Lawrence Berkeley National Laboratory, 1 Cyclotron Rd, Berkeley CA 94720-8153}

\affiliation[3]{Massachusetts Institute of Technology, 77 Massachusetts Avenue, Cambridge, MA 02139-4307}

\affiliation[4]{Georgia Institute of Technology, 801 Ferst Drive, Atlanta, GA 30332-0315}

\abstract{Large-scale optical neutrino and dark-matter detectors rely on large-area photomultiplier tubes (PMTs) for cost-effective light detection. The new R14688-100 8-inch PMT developed by Hamamatsu provides state-of-the-art timing resolution of around 1~ns (FWHM), which can help improve vertex reconstruction and enable Cherenkov and scintillation light separation in scintillation-based detectors. This PMT also provides excellent charge resolution, allowing for precision photoelectron counting and improved energy reconstruction. The \eos{} experiment is the first large-scale optical detector to utilize these PMTs. In this manuscript we present a characterization of the R14688-100 single photoelectron response, such as the transit-time spreads, the dark-rates, and the afterpulsing. The single photoelectron response measurements are performed for the 206 PMTs that will be used in \eos{}.}

\keywords{Photomultiplier tube, Eos, neutrino detectors} 

\begin{document}

\maketitle

\section{Introduction}

Large-area photomultiplier tubes (PMTs) have been used in dozens of optical neutrino and direct dark-matter experiments \cite{SNO:2002tuh,Super-Kamiokande:1998kpq,Borexino:2008dzn,DEAP:2019yzn,DayaBay:2012fng} as an effective and cost efficient way of effectively covering large surface areas with photo-sensitive detectors. An understanding of the availability and performance of different large-area PMTs is crucial for future experiments, such as \textsc{Theia} \cite{Theia:2019non}, in the selection of sensors \cite{Wen:2019sik}, as well as for accurately simulating the PMT response in existing detectors. The R14688-100 is one such large-area PMT, that will be used for the first time in a large-scale experiment in the \eos{} detector \cite{Anderson:2022lbb}.

\eos{} is a 20-ton detector, designed to perform a demonstration of key technology for future advanced neutrino detectors \cite{Anderson:2022lbb} that plan to distinguish Cherenkov and scintillation light. This technology includes spectral sorting of photons \cite{Kaptanoglu:2019gtg}, novel types of liquid scintillators \cite{YehWbLS,Biller:2020uoi}, and fast timing photodetectors \cite{Kaptanoglu:2017jxo}. The R14688-100 PMT from Hamamatsu was primarily selected for use in the \eos{} experiment due to its excellent timing resolution of around 1~ns FWHM \cite{r14688}. This can be compared to other PMTs of similar size that have timing resolutions around 2 to 3~ns \cite{r14688,Brack:2012ig,Lei:2015lua,Barros:2015pjt}. This fast timing is an important aspect in Cherenkov and scintillation separation, and it also helps to improve vertex reconstruction in optical photon-based large-scale neutrino detectors. 

A total of 206 of these PMTs were purchased to be used in \eos. The experiment requires precise characterization of each PMT to verify they each meet the detector standard in a variety of metrics. In this manuscript, we characterize the properties of these 206 PMTs. In detectors the size of \eos{} and larger, the PMTs primarily detect single photons within a given event window, and thus the single photoelectron (SPE) response is critical to determine. We also investigate the dark-rate of several PMTs, the afterpulsing rate of a single PMT, and present measurements of the PMT radio purity. These measured characteristics are additionally used in precision modeling of \eos{} in the RAT-PAC simulation software \cite{ratpac}. 

\section{R14688-100 PMT}

\begin{figure}[ht!]
    \centering
    \includegraphics[width=0.6\columnwidth]{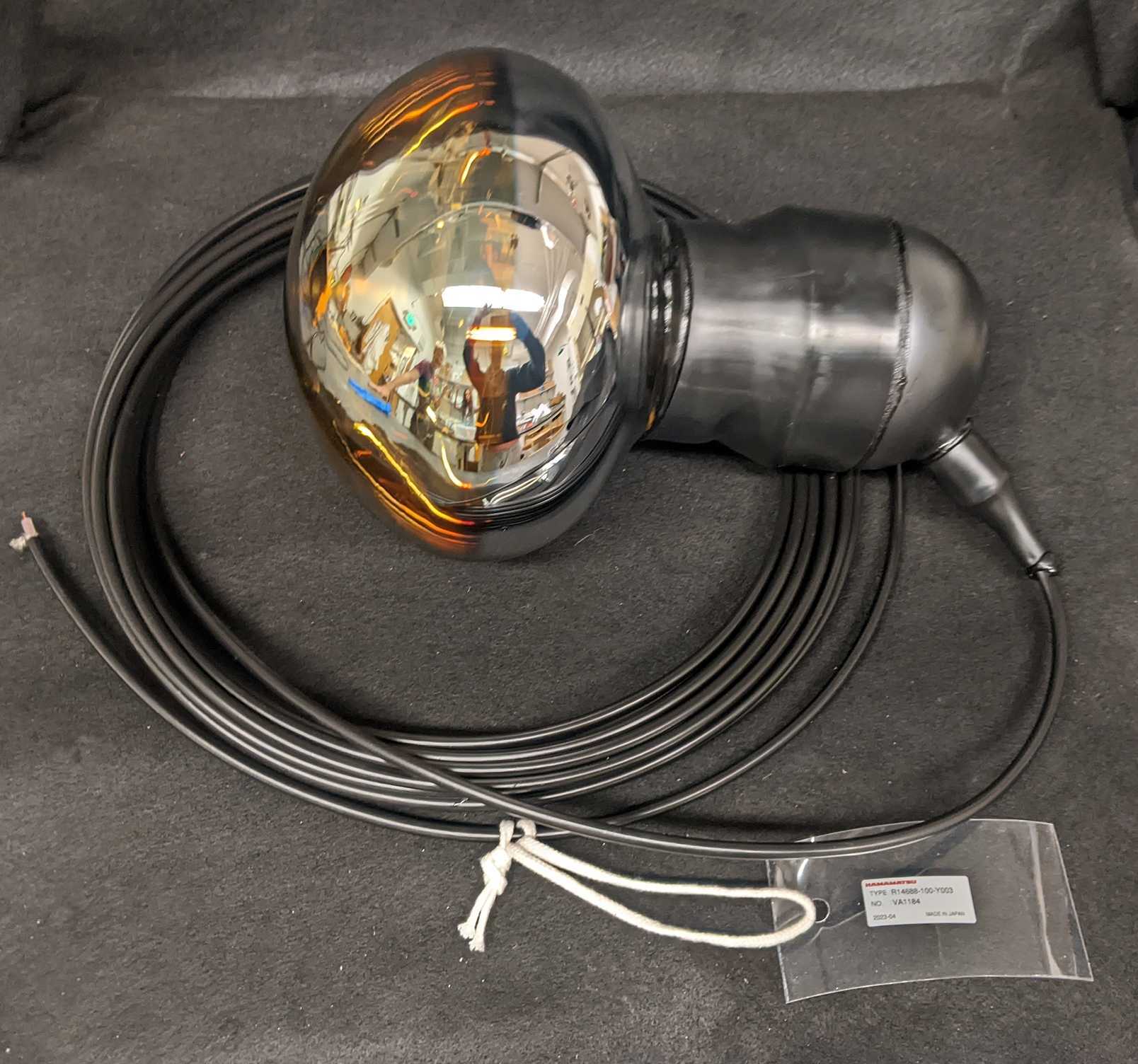}
    \caption{The R14688-100 PMT with the black water-proof potting and the 20~meter SHV cable.}
    \label{fig:pmt-pic}
\end{figure}

The R14688-100 Hamamatsu PMT is 202~mm in diameter with a high quantum efficiency, super-bialkali photocathode and an expected transit time spread (TTS) of around 1~ns full width at half maximum (FWHM) \cite{r14688}. The neck and base of the PMTs are by default water-proof potted and cabled by Hamamatsu (20~meter water proof cabling). The dynode is 10 stages and is box and linear-focused. A picture of the PMT is shown in Figure \ref{fig:pmt-pic}.

\section{SPE Characterization}\label{sec:spe-characterization}

\subsection{Experimental Setup}\label{sec:spe-experimental-setup}

The experimental setup to measure the SPE response of the PMTs consists of a central UV-transparent, cylindrical acrylic vessel (AV) which is 3~cm tall and 3~cm in diameter. A $^{90}$Sr $\beta$ source is placed above the acrylic vessel. The $\beta$ particles enter the acrylic and produce Cherenkov light. A Hamamatsu H11934-200 1-inch square PMT is optically coupled to the acrylic vessel using Eljin Technology EJ-550 optical grease. 

Two R14688-100 `measurement' PMTs are placed 20~cm on either side of the AV. The distance is selected to limit the coincidence rate between the trigger PMT and the measurement PMTs to below 5\%, which ensures we are primarily detecting SPEs. The PMTs are wrapped in a one layer of magnetic shielding that spans from the base of the PMT (where the cable exits) to the equator. The magnetic shielding protects against the Earth's magnetic field, which can deflect the photoelectrons as they travel inside of the PMT, affecting the overall collection efficiency. The strength of the magnetic field in the darkbox is approximately 50~$\mu$T, but is reduced by about a factor of 5 inside of the magnetic shielding.

\begin{figure*}[ht!]
    \centering
    \includegraphics[width=1.0\columnwidth]{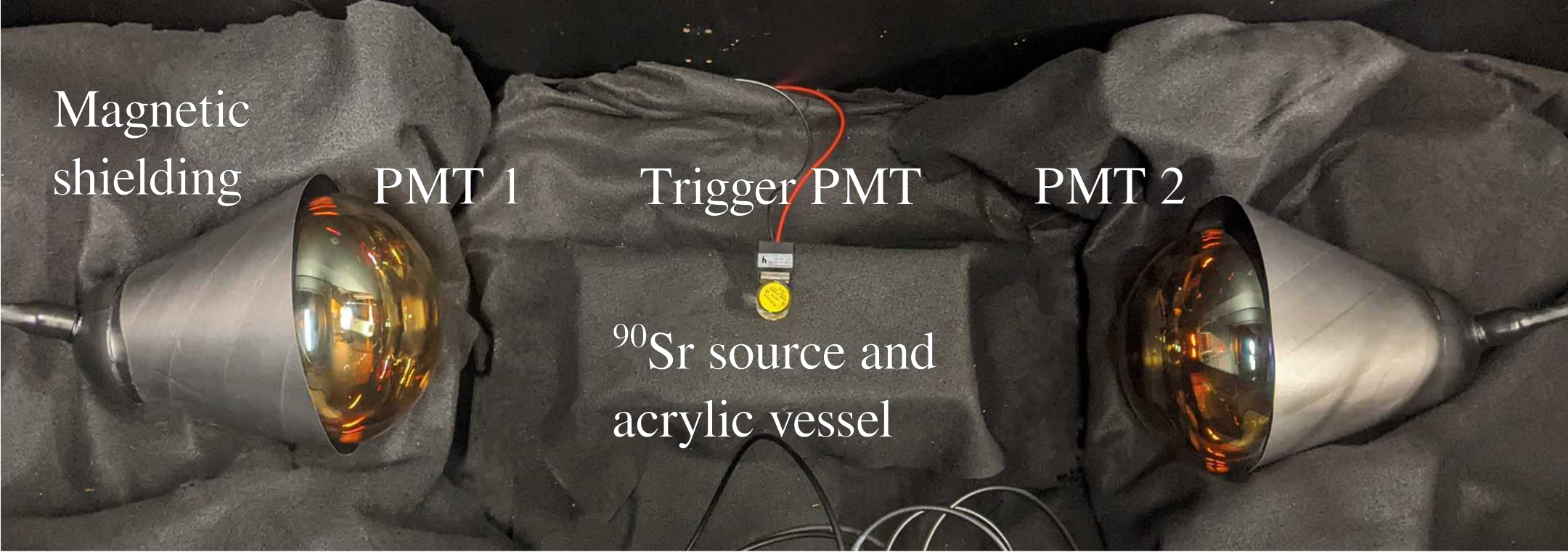}
    \caption{The experimental setup to measure the SPE response of the R14688-100 PMTs.}
    \label{fig:spe-setup}
\end{figure*}

A picture of the SPE testing setup is shown in Figure \ref{fig:spe-setup}. The space limitation of the darkbox limits the testing to two PMTs at a time. The high voltage (HV) for each PMT is selected such that the gain of the PMT is \gain. All measurements are performed at room temperature. 

\begin{figure}[ht!]
    \centering
    \includegraphics[width=0.8\textwidth]{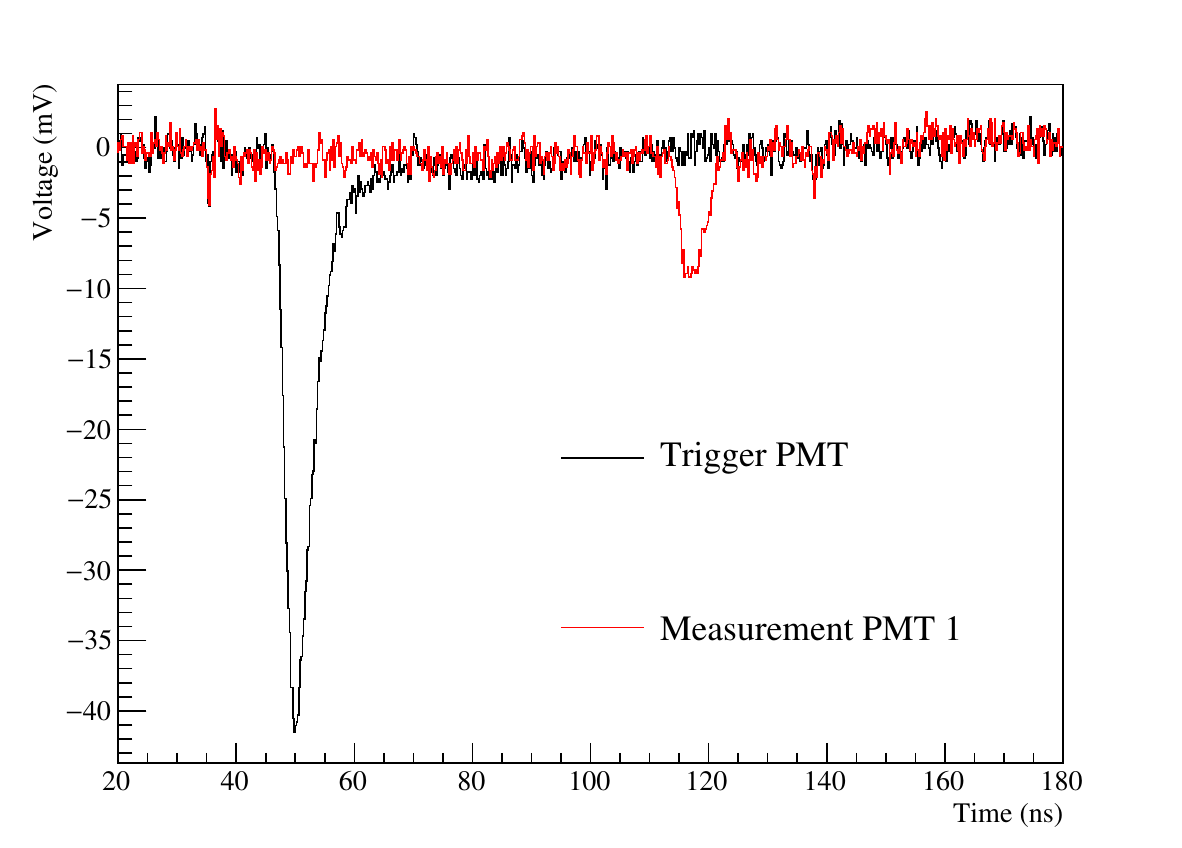}
    \caption{An example waveform consisting of a coincidence between the trigger PMT and one of the measurement PMTs (with serial number VA1206). The offset between the PMT pulses is due to cable delays.}
    \label{fig:example-waveforms}
\end{figure}

 The trigger PMT signal is input into an oscilloscope, where a $-5$ mV threshold is applied. For signals that cross the threshold, a 2~V pulse is passed to a V1742 CAEN waveform digitizer to start the data acquisition. Custom software reads out a V1742 CAEN waveform digitizer, which digitizes the signals from all three PMTs. The sampling rate of the digitizer is 5~GHz over 1024 samples and it has 12-bit ADC over a 1~V dynamic range. None of the PMT signals are additionally amplified. The waveforms for each channel are saved to an \texttt{hdf5} files. A total of 500,000 total triggers are collected, which ensures that the statistical uncertainties on the measured timing characteristics are less than 5\%. An example waveform for a trigger event is shown in Figure \ref{fig:example-waveforms}.

\subsection{Analysis}\label{sec:waveform-analysis}

The data analysis code processes the \texttt{hdf5} files output by the DAQ. First, a per-channel baseline is calculated by averaging over a 40~ns pre-trigger window. Then, for each channel, the minimum voltage is identified (the PMT pulses are downward going) and is referred to as the `peak'. The total charge deposited, $Q$, is calculated by integrating the PMT pulse in a dynamic window that ranges [-6, 12]~ns around the peak of the pulse. This calculation is also performed for the trigger PMT. 

Prior to calculating the timing, events above a threshold of about 0.2 PE are selected by enforcing $Q > 0.3$. As discussed in Section \ref{fig:spe_charge}, this removes less than 5\% of the SPE pulses. The time of the measurement PMT pulse is calculated by applying a software-based constant fraction discriminator, linearly interpolating between the samples. The fractional threshold is set to 20\%, but the precise value selected does not impact the results. For the trigger PMT, which detects many PEs per event, the time associated with a constant 2~mV threshold crossing (again, applying inter-sample interpolation) is used. This method identifies the time of the first PE detected by the trigger PMT and is more robust to the shape fluctuations in the multi-PE (MPE) pulses.

The time-difference between the measurement PMT time and the trigger PMT time, \deltat, is used to extract the timing characteristics of the measurement PMT. The trigger PMT effectively acts to produce a time-zero that has a very small jitter of about 270~ps (FWHM) \cite{h11934_datasheet}, which add negligibly to any jitter in the timing of the R14688-100 PMTs (which we expect to be around 1~ns FWHM). 

\subsection{Results}\label{sec:spe-results}

\subsubsection{Charge}\label{sec:spe-charge}

\begin{figure}[ht!]
    \centering
    \includegraphics[width=0.8\columnwidth]
    {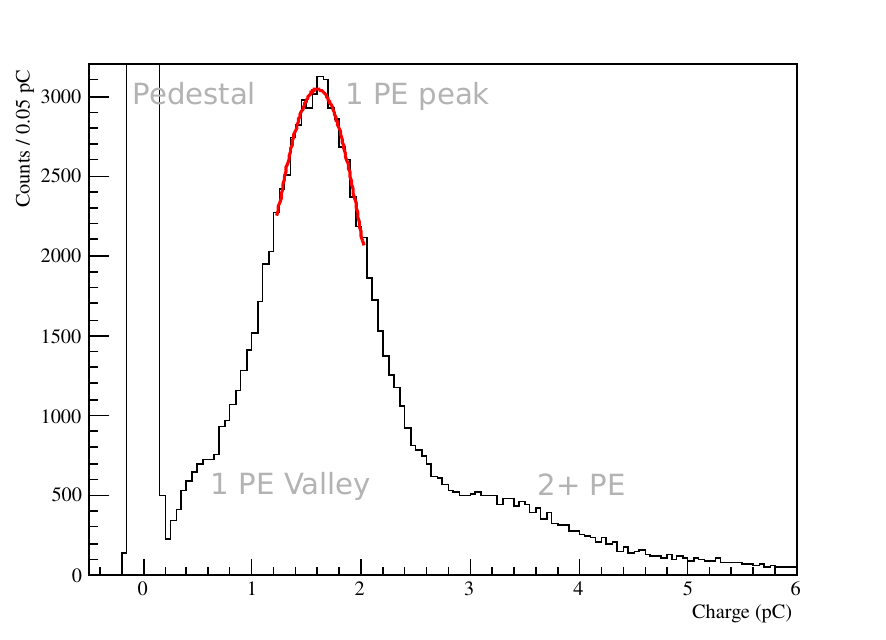}
    \caption{The charge distribution for PMT VA1206. The events around 0~pC correspond to baseline events with no photon detected. The SPE events peak around 1.6~pC and there is a small contribution for MPE events extending above 3~pC. The Gaussian fit around the SPE peak, used to extract the gain, is shown.}
    \label{fig:spe_charge}
\end{figure}

The charge distribution for the PMT with serial number VA1206 is shown in Figure \ref{fig:spe_charge}. The `pedestal' peak centered around 0~pC (which extends outside of the y-axis on the figure) corresponds to empty waveforms with no detected photon. This comprises the vast majority of the coincidences, as the coincidence rate is selected to be below 5\%. The peak around 1.6~pC corresponds to the detection of single photons with a gain of \gain. The value of the peak is determined by identifying the maximum bin above 0.2~pC and fitting a Gaussian over the range that is 0.4~pC on either side of the peak.

As identifiable in Figure~\ref{fig:spe_charge} the one PE peak is well separated from the zero PE peak. This is often quantified using a peak-to-valley (P/V) ratio. The P/V is calculating by identifying the total counts in the peak bin and dividing by the total counts in bin at 0.5~pC. The selection of this value is arbitrary, but it allows for a stable definition that is not impact by small changes in the width of the baseline peak around 0~pC. The P/V for VA1206 is 4.58, and the PMTs typically have values above 4. This can be compared directly to other types of large-area PMTs, which have values around 2 to 3 \cite{Brack:2012ig,Akindele:2023ixz}.

Because the P/V is so large for these PMTs, relatively few PMT pulses fall below the 0.3~pC threshold applied for the timing measurement, described below. If the charge distribution extrapolates linearly down to 0~pC under the pedestal, less than 5\% of the SPE pulses are lost from the roughly 0.2~PE threshold. This highlights the advantage of the large P/V ratio for these PMTs, indicating that any threshold applied below the peak will have a relatively small efficiency sacrifice.

\subsubsection{Timing}\label{sec:spe-timing}

\begin{figure}[ht!]
    \centering
    \includegraphics[width=0.8\columnwidth]{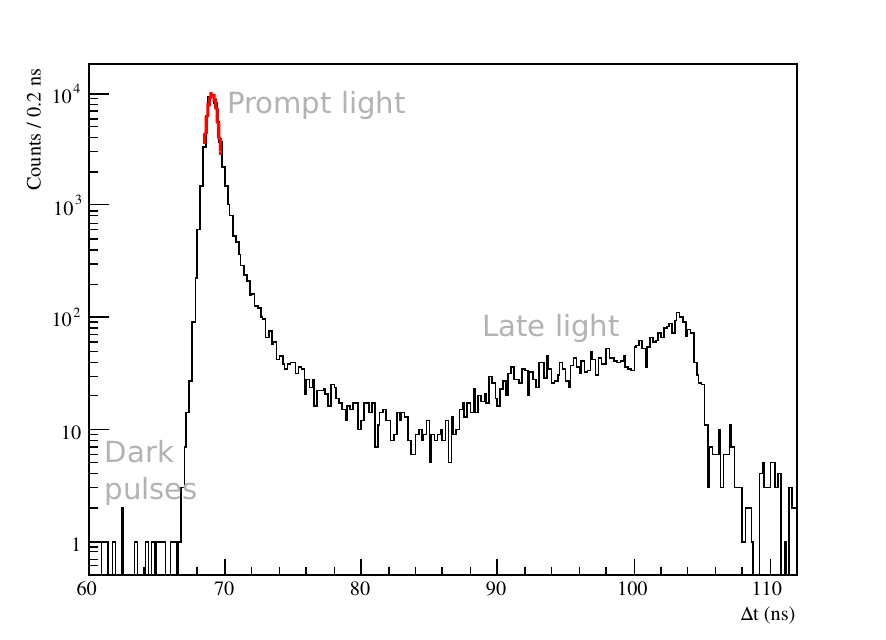}
    \caption{The \deltat{} distribution for PMT VA1206. The Gaussian fit around the prompt peak, used to extract the TTS, is shown. Note the offset in \deltat{} from zero is set by cable delays and is arbitrary.}
    \label{fig:tts}
\end{figure}

The \deltat{} distribution for the PMT with serial number VA1206, running at a gain of \gain{}, is shown in Figure \ref{fig:tts}. The overall offset of \deltat{} from zero is set by cable delays and is arbitrary. As discussed, because the trigger PMT has an extremely fast response, the \deltat{} distribution is primarily showing the transit-time (TT) distribution for single photoelectrons detected by the measurement PMT. The width of the prompt light, peaking around 70~ns in the \deltat{} distribution, can be fit with a Gaussian function to extract the transit-time spread (TTS).

Because the prompt TT distribution is not perfectly Gaussian (rather, it is slightly skewed right), the fit-range selected impacts the measured TTS distribution. After binning the \deltat{} distribution in 0.2~ns bins, the peak is identified and a Gaussian is fit using a range that is $\pm0.6$~ns around the peak. The FWHM of the fitted Gaussian is taken to be the TTS of the PMT. By adjusting the fit range between $0.2$ to $1.0$~ns in steps of $0.2$~ns, the fitted value of the TTS varies at most by 5\%, which is the largest source of systematic uncertainty. 

The values of the TTS are reported using the $\pm0.6$~ns range. The average TTS across the PMTs is 1.00~ns and the standard deviation is 0.08~ns. This indicates that these PMTs achieve a state-of-the-art timing resolution and that the behavior is very consistent from PMT to PMT. 

In addition to the prompt light, there is a clear contribution of light that comes between 5 to 40~ns late, as identified in Fig. \ref{fig:tts}. This effect is due to  elastic scatters of the PE off of the first dynode, before being accelerated back some time later. The fraction of detected photons that are late is quantified by identifying the prompt peak and integrating a window that is between 5 and 40~ns after the prompt peak and comparing that value to the integral of the full timing window. Averaging over all of the tested PMTs, the fraction of late-light is 9.39 $\pm$ 0.58\%. 

\subsubsection{High Voltage Scan}\label{sec:voltage-scan}

\begin{figure}[ht!]
    \centering
    \includegraphics[width=0.45\columnwidth]{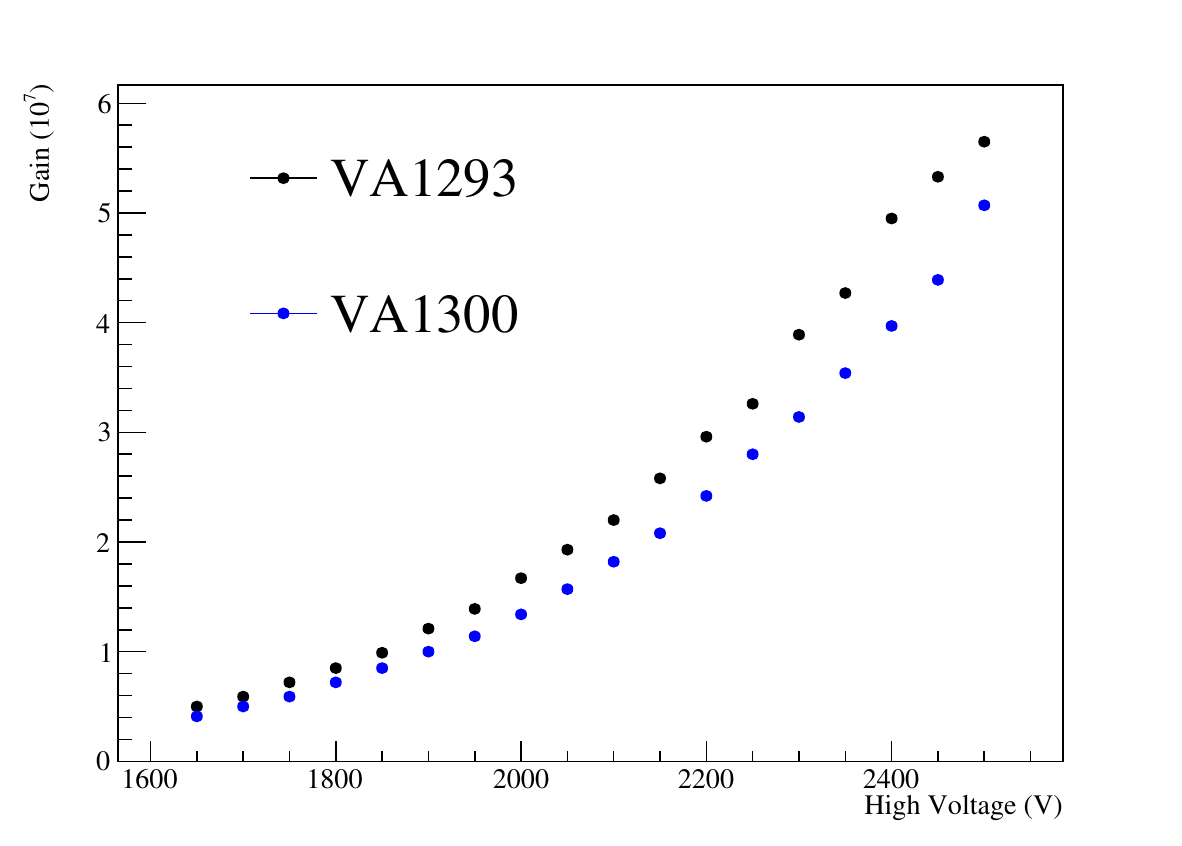}
    \includegraphics[width=0.45\columnwidth]{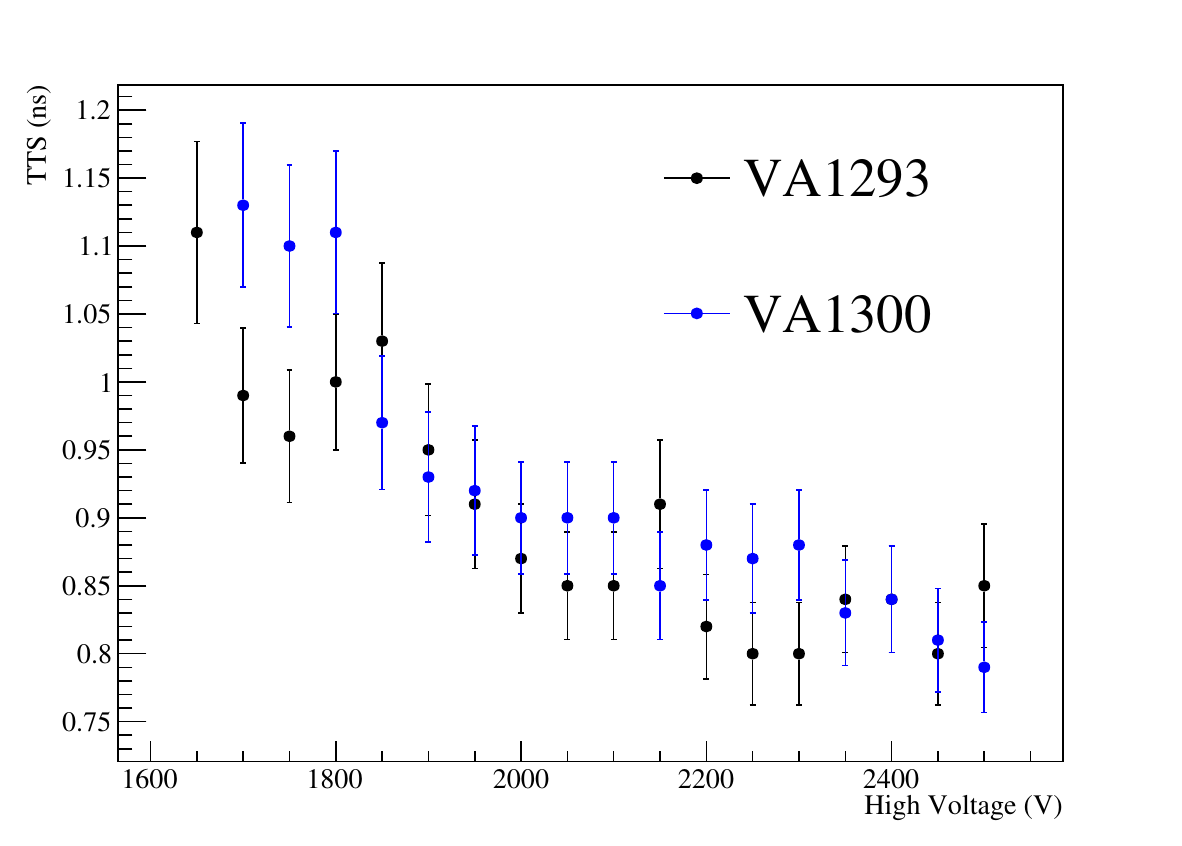}
    \caption{The gain (left) and TTS (right) as a function of the supply voltage for two PMTs. The uncertainties on the TTS include the statistical and systematic errors summed in quadrature.}
    \label{fig:scan}
\end{figure}

In order to determine the high voltage that yields a gain of \gain{} the high voltage is scanned across a range of values, and the gain is determined by measuring the mean of the charge distribution. For two PMTS, VA1293 and VA1300, we investigated the TTS and gain as a function of high voltage from 1650~V to 2500~V. The results are shown in Figure \ref{fig:scan}. Notably, at large gains of around 5$\times 10^{7}$, reached with supply voltages around 2400 to 2500~V, the TTS is improved from around 1~ns to about 0.8~ns. Although large-area PMTs are often operated at gains of $10^{7}$ for reasons such as long-term stability, running these PMTs at higher gains would yield TTSs smaller than 1~ns.

\subsubsection{Dark-Rate}\label{sec:dark-rate}

The constrained timeline for the \eos{} construction and the installation of the PMTs restrained the total number of PMTs that could be dark-rate tested, but 14 PMTs were successfully tested. The dark-rate of the PMTs is measured at least twenty hours after the dark-box has been closed, allowing the PMT dark-rate to settle to baseline.

The rate is extracted by fitting a 50~ns pre-trigger window, prior to the Cherenkov light emission, with a flat polynomial. The total counts in that window is converted to a rate and is well-described by the uniform model. Over the 14 PMTs measured we find an average dark-rate of 2175 $\pm$ 1400 Hz, with a minimum rate of 590 Hz and a maximum rate of 5024 Hz. This level of variation is typical for large-area PMTs \cite{Brack:2012ig,Akindele:2023ixz}.

The measurements are performed at room temperature around 22$^{\circ}$~C. The expected temperature of the water in \eos{} is 15$^{\circ}$~C, which would lead to an approximate decrease by a factor of 2-3 for the dark-rates \cite{handbook,Akindele:2023ixz}. The measured dark-rates for these PMTs are similar to those reported for other large-area PMTs \cite{r14688,Brack:2012ig,Lei:2015lua,Barros:2015pjt,Kaptanoglu:2017jxo}.

\subsubsection{SPE summary}\label{sec:spe-summary}

A summary of the SPE testing results is provided in Table \ref{tab:spe_results}. The SPE results for every PMT is provided in \ref{sec:all-pmt-data}.

\begin{table}[ht!]
    \centering
    \begin{tabular}{l|c|c|c} \hline \hline
         Parameter & Mean & Maximum & Minimum \\ \hline 
         HV (V) & 1989 $\pm$ 143 & 2500 & 1700 \\ 
         Charge P/V & 5.69 $\pm$ 0.92 & 7.36 & 3.97 \\
         TTS (ns) & 1.00 $\pm$ 0.08 & 1.23 & 0.83 \\
         Late-Fraction (\%) & 9.39 $\pm$ 0.58 & 11.2 & 7.46 \\ \hline 
    \end{tabular}
    \caption{The mean, standard deviation, maximum, and minimum values of the measured SPE parameters for a gain of \gain.}
    \label{tab:spe_results}
\end{table}

\section{Afterpulsing}

Signal generated by residual ions inside of the PMT can cause delayed signals that occur hundreds of ns or several $\mu$s after an initial PE. Specifically, a generated PE can ionize residual gases inside of the PMT, such as helium. The ions then slowly drift back to the photocathode, where they are absorbed, creating a single PE or multiple PEs that are delayed relative to the initial signal. The time-difference between the initial signal and the delayed signal is determined by the type of gas in the PMT, the electric field strength, and the size of the PMT. Notably, these afterpulses are produced via a different mechanism then what we call late-pulses in Section \ref{sec:spe-timing}. For reference, the formation of afterpulses in large-area PMTs has been studied in Refs.~\cite{Kaptanoglu:2017jxo,AKCHURIN2007121,Ma_2011}.

\subsection{Experimental Setup}

A blue LED is directed at an R14688-100 PMT, operating at a gain of $10^{7}$, and pulsed at 100 Hz with a 30~ns pulse width. The amplitude of the pulse is adjusted to change the total amount of prompt PE detected. Data is collected across a variety of different intensities to study the afterpulsing rate as a function of the number of PE in the prompt pulse.

The CAEN V1742 digitizer has a maximum window width of about 2~$\mu$s, which was insufficient for measuring the afterpulses. Therefore, the afterpulsing setup digitized the PMT waveforms on a Lecroy Waverunner 9054 at 1 GS/s over a 50~$\mu$s window. The data is collected using custom software called \texttt{lecrunch} which outputs the waveforms to \texttt{hdf5} files.

\subsection{Analysis}

In order to identify the prompt pulse, generated by the LED light, and the subsequent afterpulses, the waveform is split into distinct windows. First, a 0.2~$\mu$s window at the beginning of the waveform is used to calculate the baseline. Then a 19.6~$\mu$s pre-trigger window is used to identify dark-pulses, in order to correct for their contribution in the afterpulsing window. A 0.4~$\mu$s window is used to identify the location of the prompt pulse and the following 29.8~$\mu$s is used to look for afterpulsing.

In the primary peak, where we expect the data to be MPE, the pulse is easily identified and integrated over a 60~ns window (centered around the peak). To roughly calculate the number of PE (NPE) detected in this window, the integrated charge is divided by the value of mean of the SPE charge distribution. The afterpulsing rate is then measured as a function of the NPE in the primary pulse. 

The pulses in the dark-window and the afterpulsing window are identified by applying a 3~mV threshold and identifying threshold crossings. A threshold counting is only counted if the waveforms is below threshold on the previous sample. This threshold corresponds to roughly 0.2~PE.

\subsection{Results}

Five datasets are collected, varying the intensity of the LED. In the first dataset, the LED is dimly lit, with an average occupancy of around 2~PE. In the final dataset, the LED is pulsed such that the PMT has an average occupancy of almost 30~PE per pulse.

\begin{figure}[ht!]
    \centering
    \includegraphics[width=0.8\columnwidth]{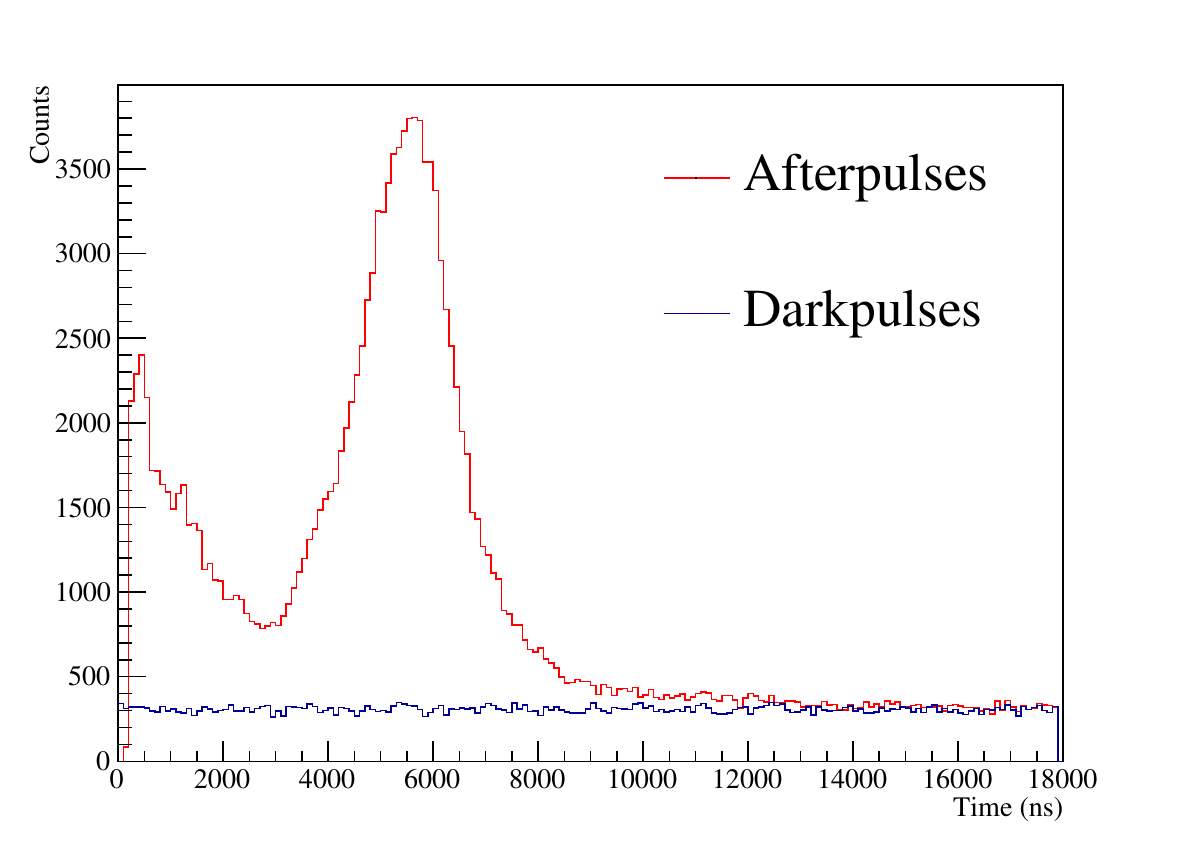}
    \caption{The timing distribution of the afterpulses for a dataset with an average occupancy of 5~PE. The events from the dark-rate, taken from a pre-trigger window, are shown as a direct comparison. The prompt light arrives at 0~ns for this figure.}
    \label{fig:ap_timing}
\end{figure}

The time distribution of the afterpulses for one of the datasets, with an average occupancy of 5~PE, is shown in Figure \ref{fig:ap_timing}. The afterpulsing distribution shows two clear peaks around 500~ns and 5.0~$\mu$s after the initial prompt light from the LED, which occurs at 0~ns in Figure \ref{fig:ap_timing}. The locations and heights of each of these peaks reflects different gases at different amounts present in the PMT. By around 16.0$\mu$s we no longer identify any afterpulsing above the dark noise background. The sharp cut-off at shorter times is due to the 400~ns window used to identify the hit-time of the prompt light, and it is possible some of the afterpulses with shorter times are not identified.

\begin{figure}[ht!]
    \centering
    \includegraphics[width=0.8\columnwidth]{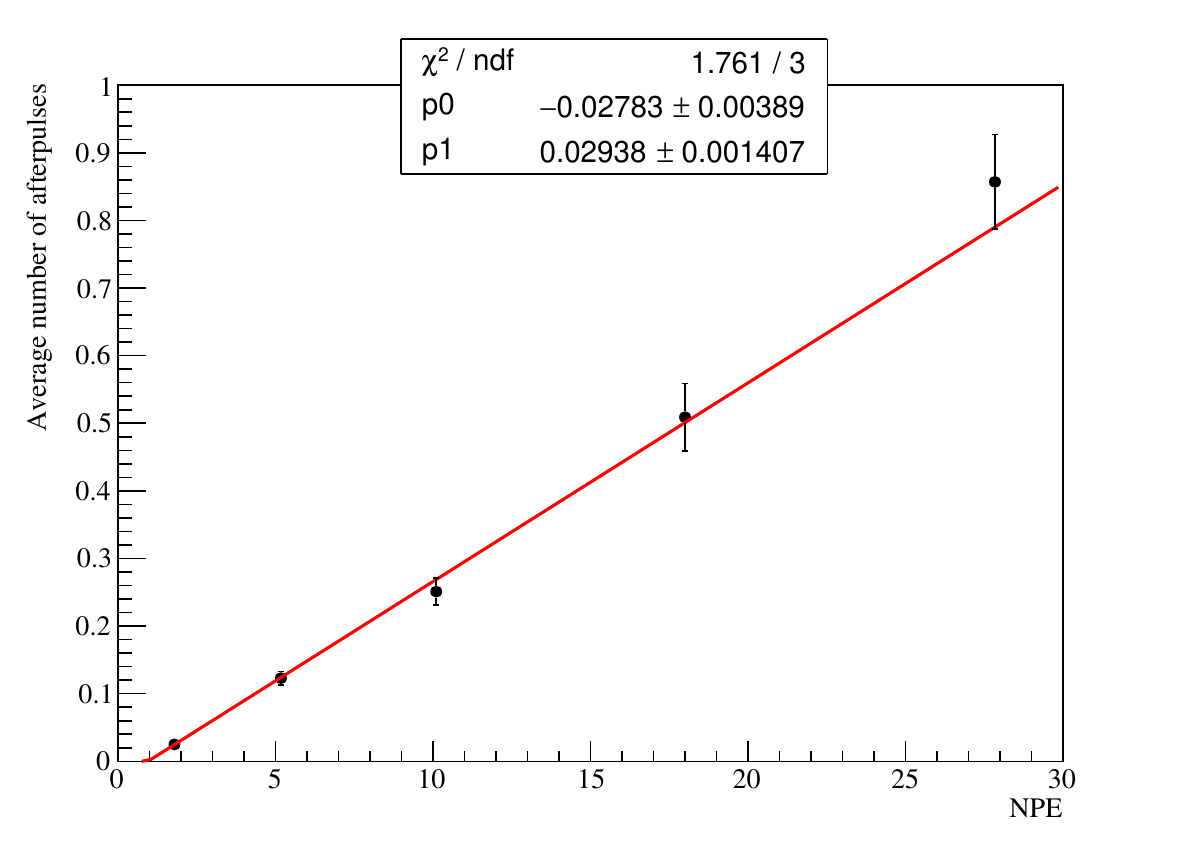}
    \caption{The average number of afterpulses detected for a given number of prompt NPE. A linear fit indicates that the average number of afterpulses produced for a 1~PE signal is 0.016.}
    \label{fig:ap_rate}
\end{figure}

Most importantly, we study the afterpulsing rate as a function of the number of detected PE from the prompt LED light. These results are given in Figure \ref{fig:ap_rate}. The figure shows the average number of afterpulses as a function of NPE, where a value of 0.5 indicates that in half of the events there was an afterpulse identified. A value larger than one would indicate more than one afterpulse per prompt LED pulse.  

For the dimmest LED setting, with an integrated charge that indicates that we are primarily measuring one or two PE per LED pulse, the afterpulsing rate was about 2.5\%. The linear fit indicates an average rate of 1.6\% for an occupancy of 1PE; however, the fit clearly breaks down at very low occupancy, indicated by the negative value for the y-intercept. The linear fit is useful for interpolating and extrapolating to higher occupancy. 

It should be noted that the calculation of the average number of afterpulses does not include several factors. First, no correction is made for an efficiency loss from the SPE selection, and a rough efficiency loss of 5\% was estimated in Section~\ref{sec:spe-charge}. Second, any afterpulses with timing less than 400~ns are not included in this calculation, as discussed in relation to Figure \ref{fig:ap_timing}. Again, we expect this to have a relatively small effect, as the 400~ns time-window is short relative to the 20~$\mu$s window to identify afterpulses. Finally, we do not account for MPE afterpulses, and assume the pulses are SPE. To check this assumption, we look at the charge distribution of the afterpulses, and even at the highest LED intensity, we find only about 5\% of the afterpulses are MPE. 

\section{Radiopurity Assay}

PMTs are often a source of significant radioactive backgrounds in neutrino and dark-matter experiments \cite{Akerib:2012da}. This comes primarily from uranium and thorium intrinsic to the PMT glass and dynode that produces $\gamma$-rays that can travel through buffer material into the sensitive detector volume.

Samples taken from an intentionally broken R14688-100 PMT, provided by Hamamatsu, were sent to the low background counting facilities at the Black Hills Underground Campus. Measurements were performed on the 4850’ level of the Sanford Underground Research Facility (SURF) \cite{surf_counting, LZ:2020fty, LZ:2022ysc} to measure the radioactivity of the PMT. The samples were broadly grouped in two categories: (1) the crushed glass from the PMT bulb and (2) the metallic and ceramic parts the make-up the dynode structure and the rest of the PMT. These two groups are measured separately in the SURF facilities using the Morgan detector station (see Ref. \cite{surf_counting} for details). The results are provided in Table \ref{tab:radioactivity}. The total weight of the glass and dynode parts is 0.51 kg and 0.39 kg respectively. Notably, Hamamatsu communicated that the glass for these PMTs is possible to construct with lower radioactivity \cite{private_comm}. 

\begin{table}[ht!]
    \centering
    \begin{tabular}{c|c|c|c} \hline \hline 
         Sample & U (mBq/kg) & Th (mBq/kg) & K (mBq/kg) \\ \hline 
         PMT glass & 3144 $\pm$ 158 & 4486 $\pm$ 257 & 7874 $\pm$ 398 \\ 
         PMT dynode & 421.2 $\pm$ 28.2 & 55.8 $\pm$ 5.0 & 219 $\pm$ 24 \\ \hline 
    \end{tabular}
    \caption{The results of the radiological assay of the PMT parts, performed at SURF.}
    \label{tab:radioactivity}
\end{table}

\section{Discussion}

Based on the results shown in this manuscript, the R14688-100 is a promising candidate for future large-scale neutrino experiments, such as Theia, that hope to utilize fast timing to improve background rejection \cite{Theia:2019non}. Additionally, existing detectors with planned upgrades that already utilize 8-inch PMTs, such as SNO+ \cite{SNO:2021xpa}, could directly replaced old PMTs with these modern PMTs. This improvement would more than double the detection efficiency and improve the TTS by more than a factor of three \cite{BILLER1999364}.

Additionally, the ability to identify the number of photons detected by a PMT will be important for future scintillation based neutrino detectors \cite{Akashi-Ronquest:2014jga}. This would help improve energy resolution by providing a more precise measure of the total light detected. Thus, the ability to distinguish one PE from zero, two PE from one PE, etc. will be an important aspect of future photosensors. Given the large P/V ratio measured in this manuscript, the R14688-100 PMT provides excellent discrimination capability for PE counting. 

The timing of the R14688-100 PMT, measured to be on average around 1~ns (FWHM) across 206 PMTs, is the primary motivation for using this large-area PMT. This state-of-the-art timing is at least twice as good as other large-area PMTs, and could significantly improve reconstruction performance and background rejection in future scintillation-based neutrino experiments \cite{Land:2020oiz}. About 10\% of the light is detected late relative to the prompt peak, which is typical for large-area PMTs. The dark-rate of the PMTs has not been robustly characterized at the temperatures standard for large detectors; however, initial measurements of several PMTs at room temperature indicate dark-rates consistent with other PMTs of similar sizes. Lastly, we performed an afterpulsing measurement of a single PMT, which indicated afterpulsing rates around 2\% per PE, with peaks in the timing distribution around 500~ns and 5~us. 

\section{Conclusion}\label{sec:conclusion}

In this manuscript the SPE detection performance is characterized for 206 large-area Hamamatsu R14688-100 PMTs. The fast timing and excellent P/V makes it an ideal PMT for use in \eos{}, where these features will be used to perform Cherenkov and scintillation separation and precision photon counting. Ultimately, this will help improve reconstruction and reduce backgrounds. These measurements can also be directly input into simulation software in order to better simulate the detector response. Finally, measurements made of  afterpulsing rate, dark-rate, and radioactivity of the PMT provide important details for \eos{} and for future experiments deciding on the type of PMT to use. 

\section*{Data Availability Statement}

The data used in this manuscript is available in an accessible \texttt{postgres} database hosted at Lawrence Berkeley National Lab. If interested, please contact the author for database connection information. The results for the full list of all tests performed on the Eos PMT is available at \url{https://nino.lbl.gov/pmts/}.

\section{Acknowledgements}

Thank you kindly to Brianna Mount and Douglas Tiedt for performing the background counting measurements and analysis at the Black Hills State University Underground Campus and the Berkeley Low Background Facility. The work conducted at Lawrence Berkeley National Laboratory was performed under the auspices of the U.S. Department of Energy under Contract DE-AC02-05CH11231. This project was funded by the U.S. Department of Energy, National Nuclear Security Administration, Office of Defense Nuclear Nonproliferation Research and Development (DNN R\&D). Author M. Duce was supported by the Department of Energy and National Nuclear Security Administration, United States through the Enabling Technology and Innovation Consortium under Award Number DE-NA0003921.

\clearpage

\appendix

\section{List of Results for All PMTs}\label{sec:all-pmt-data}

Table \ref{tab:all_spe_results} provides a list of the high voltage, charge peak, charge peak-to-valley, TTS, and late-pulsing percentage for all PMTs tested in the SPE setup.

\begin{longtable}{c|c|c|c|c|c} \hline \hline 
    PMT ID & High Voltage & $Q$ Peak [pC] & $Q$ P/V & TTS [ns] & Late-light Pct. (\%) \\ \hline 
VA1059 & 1960 & 1.54 & 5.62 & 1.035 & 8.81 \\
VA1060 & 2395 & 1.62 & 5.29 & 0.835 & 8.83 \\
VA1062 & 1980 & 1.54 & 5.53 & 0.915 & 8.98 \\
VA1064 & 1865 & 1.58 & 5.75 & 0.985 & 8.96 \\
VA1065 & 1950 & 1.62 & 4.87 & 1.008 & 9.27 \\
VA1066 & 1875 & 1.57 & 5.36 & 1.053 & 9.82 \\
VA1067 & 1935 & 1.57 & 4.76 & 1.038 & 9.64 \\
VA1068 & 1720 & 1.58 & 4.32 & 1.115 & 9.18 \\
VA1070 & 1900 & 1.52 & 4.59 & 1.226 & 8.18 \\
VA1071 & 1900 & 1.59 & 3.97 & 1.113 & 9.03 \\
VA1072 & 1830 & 1.61 & 5.24 & 0.975 & 9.17 \\
VA1073 & 1950 & 1.59 & 4.68 & 1.014 & 9.18 \\
VA1074 & 1900 & 1.56 & 5.85 & 1.056 & 9.79 \\
VA1075 & 1920 & 1.49 & 6.84 & 1.020 & 8.89 \\
VA1076 & 1900 & 1.62 & 5.89 & 1.091 & 9.45 \\
VA1077 & 1850 & 1.54 & 4.73 & 1.137 & 8.51 \\
VA1078 & 1800 & 1.57 & 5.71 & 1.137 & 8.76 \\
VA1080 & 1845 & 1.56 & 5.11 & 1.044 & 8.58 \\
VA1081 & 1700 & 1.58 & 5.80 & 1.108 & 8.12 \\
VA1082 & 1705 & 1.58 & 5.24 & 1.107 & 8.43 \\
VA1084 & 1940 & 1.56 & 6.73 & 1.041 & 8.93 \\
VA1086 & 1955 & 1.56 & 6.08 & 1.017 & 8.84 \\
VA1088 & 2040 & 1.63 & 6.48 & 0.932 & 9.88 \\
VA1089 & 1790 & 1.58 & 5.64 & 1.101 & 9.51 \\
VA1090 & 1840 & 1.56 & 6.05 & 1.165 & 8.73 \\
VA1091 & 1820 & 1.57 & 7.15 & 1.113 & 8.73 \\
VA1092 & 1810 & 1.59 & 6.33 & 1.045 & 8.16 \\
VA1093 & 1745 & 1.57 & 6.09 & 1.188 & 9.20 \\
VA1094 & 1725 & 1.62 & 6.09 & 1.149 & 9.44 \\
VA1095 & 1720 & 1.61 & 4.96 & 1.011 & 8.86 \\
VA1097 & 2020 & 1.68 & 6.98 & 0.976 & 9.17 \\
VA1098 & 2105 & 1.57 & 5.68 & 0.987 & 9.10 \\
VA1099 & 1980 & 1.58 & 6.58 & 0.928 & 9.82 \\
VA1100 & 2100 & 1.55 & 4.94 & 0.958 & 9.19 \\
VA1101 & 1860 & 1.62 & 5.42 & 1.134 & 9.42 \\
VA1102 & 1925 & 1.60 & 6.07 & 1.087 & 10.14 \\
VA1103 & 1950 & 1.58 & 6.39 & 1.030 & 9.35 \\
VA1104 & 2080 & 1.61 & 5.56 & 0.946 & 9.94 \\
VA1105 & 2050 & 1.56 & 5.77 & 0.942 & 9.62 \\
VA1106 & 1955 & 1.59 & 5.70 & 0.975 & 10.05 \\
VA1107 & 1895 & 1.52 & 5.98 & 1.011 & 9.86 \\
VA1108 & 2020 & 1.59 & 5.64 & 1.004 & 9.97 \\
VA1110 & 1980 & 1.55 & 6.26 & 0.988 & 10.38 \\
VA1111 & 2150 & 1.56 & 6.95 & 1.019 & 9.55 \\
VA1113 & 1840 & 1.60 & 6.82 & 1.043 & 9.98 \\
VA1114 & 1880 & 1.58 & 6.36 & 1.041 & 9.86 \\
VA1115 & 1870 & 1.58 & 6.47 & 1.030 & 9.91 \\
VA1116 & 2070 & 1.59 & 5.62 & 0.970 & 10.01 \\
VA1117 & 2085 & 1.63 & 5.76 & 0.992 & 9.87 \\
VA1119 & 1825 & 1.63 & 5.48 & 1.114 & 9.60 \\
VA1120 & 2110 & 1.55 & 5.71 & 0.885 & 9.65 \\
VA1121 & 1915 & 1.61 & 5.81 & 1.057 & 9.67 \\
VA1123 & 1845 & 1.62 & 7.16 & 1.020 & 9.86 \\
VA1124 & 1790 & 1.61 & 6.90 & 1.106 & 9.64 \\
VA1125 & 1900 & 1.66 & 6.09 & 0.974 & 10.34 \\
VA1126 & 1935 & 1.56 & 5.80 & 1.109 & 9.53 \\
VA1127 & 1870 & 1.56 & 6.62 & 1.039 & 9.64 \\
VA1128 & 1955 & 1.62 & 5.53 & 1.001 & 9.65 \\
VA1129 & 1915 & 1.55 & 6.76 & 1.118 & 9.94 \\
VA1130 & 1865 & 1.56 & 7.14 & 1.060 & 9.82 \\
VA1131 & 1945 & 1.60 & 6.31 & 0.975 & 10.10 \\
VA1132 & 1825 & 1.56 & 6.13 & 1.087 & 9.16 \\
VA1134 & 2030 & 1.56 & 5.74 & 0.970 & 10.38 \\
VA1135 & 1860 & 1.63 & 6.95 & 1.008 & 9.77 \\
VA1136 & 2010 & 1.62 & 5.79 & 1.095 & 10.93 \\
VA1137 & 2025 & 1.73 & 6.59 & 1.072 & 11.17 \\
VA1138 & 2100 & 1.69 & 6.18 & 1.005 & 10.15 \\
VA1139 & 1955 & 1.69 & 6.18 & 1.005 & 10.15 \\
VA1140 & 2100 & 1.60 & 5.76 & 0.977 & 9.83 \\
VA1141 & 1945 & 1.63 & 6.48 & 1.002 & 9.99 \\
VA1142 & 2160 & 1.54 & 5.78 & 0.864 & 9.46 \\
VA1143 & 2220 & 1.80 & 5.40 & 1.011 & 10.29 \\
VA1144 & 1930 & 1.55 & 5.09 & 1.105 & 9.40 \\
VA1145 & 2065 & 1.62 & 5.74 & 1.000 & 10.66 \\
VA1146 & 1995 & 1.55 & 5.80 & 1.018 & 9.79 \\
VA1148 & 1980 & 1.56 & 5.53 & 0.997 & 10.08 \\
VA1149 & 2045 & 1.52 & 6.03 & 1.010 & 10.24 \\
VA1150 & 1870 & 1.60 & 5.91 & 1.033 & 9.64 \\
VA1151 & 1930 & 1.51 & 5.67 & 1.164 & 9.61 \\
VA1252 & 2095 & 1.53 & 6.12 & 0.908 & 9.77 \\
VA1154 & 1990 & 1.59 & 6.31 & 0.983 & 9.37 \\
VA1155 & 2050 & 1.53 & 6.14 & 1.101 & 9.40 \\
VA1156 & 2015 & 1.59 & 6.23 & 0.947 & 9.95 \\
VA1157 & 2170 & 1.55 & 5.03 & 0.897 & 9.43 \\
VA1158 & 2150 & 1.55 & 5.96 & 0.946 & 9.71 \\
VA1159 & 1920 & 1.58 & 6.29 & 1.072 & 9.06 \\
VA1160 & 1900 & 1.56 & 5.73 & 1.059 & 9.26 \\
VA1161 & 1905 & 1.54 & 5.37 & 0.961 & 9.42 \\
VA1162 & 2150 & 1.66 & 5.26 & 0.834 & 10.27 \\
VA1164 & 2130 & 1.64 & 5.70 & 0.938 & 9.54 \\
VA1165 & 2035 & 1.61 & 6.22 & 0.978 & 8.60 \\
VA1166 & 2190 & 1.54 & 6.02 & 0.841 & 9.65 \\
VA1168 & 1900 & 1.59 & 7.26 & 1.035 & 9.18 \\
VA1170 & 2265 & 1.65 & 6.57 & 0.921 & 9.74 \\
VA1172 & 2280 & 1.52 & 4.24 & 0.911 & 8.56 \\
VA1173 & 2200 & 1.52 & 5.51 & 0.953 & 9.29 \\
VA1174 & 2060 & 1.60 & 6.52 & 0.921 & 8.91 \\
VA1175 & 2060 & 1.63 & 4.43 & 1.026 & 7.46 \\
VA1176 & 2150 & 1.63 & 6.20 & 0.853 & 9.77 \\
VA1177 & 2040 & 1.61 & 7.27 & 0.919 & 9.61 \\
VA1178 & 2040 & 1.48 & 5.37 & 0.922 & 9.40 \\
VA1179 & 2040 & 1.68 & 5.42 & 0.987 & 9.72 \\
VA1180 & 2255 & 1.62 & 5.30 & 0.939 & 9.60 \\
VA1181 & 2180 & 1.60 & 6.01 & 0.878 & 9.35 \\
VA1182 & 2350 & 1.31 & 4.60 & 0.924 & 8.78 \\
VA1183 & 2010 & 1.45 & 6.42 & 0.912 & 9.24 \\
VA1184 & 1980 & 1.52 & 5.83 & 0.980 & 9.02 \\
VA1185 & 1955 & 1.64 & 6.25 & 0.959 & 9.75 \\
VA1186 & 1995 & 1.59 & 5.88 & 0.974 & 9.27 \\
VA1187 & 1925 & 1.55 & 5.37 & 0.918 & 9.60 \\
VA1188 & 2165 & 1.50 & 5.40 & 0.929 & 9.37 \\
VA1189 & 2100 & 1.57 & 4.99 & 0.933 & 9.85 \\
VA1190 & 2050 & 1.60 & 5.75 & 1.016 & 9.40 \\
VA1191 & 2180 & 1.56 & 5.43 & 0.969 & 9.22 \\
VA1192 & 1955 & 1.53 & 6.47 & 0.927 & 9.00 \\
VA1193 & 2080 & 1.56 & 5.76 & 1.075 & 9.44 \\
VA1195 & 1720 & 1.60 & 6.46 & 1.123 & 9.14 \\
VA1196 & 2500 & 1.50 & 4.75 & 0.920 & 9.57 \\
VA1197 & 2220 & 1.52 & 5.17 & 0.968 & 8.59 \\
VA1202 & 2310 & 1.46 & 4.43 & 0.846 & 8.99 \\
VA1205 & 2260 & 1.59 & 4.44 & 0.939 & 9.11 \\
VA1206 & 2110 & 1.60 & 4.58 & 0.914 & 9.25 \\
VA1207 & 2110 & 1.78 & 5.20 & 0.895 & 9.82 \\
VA1208 & 2290 & 1.62 & 5.18 & 0.874 & 10.02 \\
VA1213 & 2205 & 1.54 & 4.54 & 0.880 & 9.41 \\
VA1215 & 1935 & 1.54 & 4.05 & 0.934 & 9.17 \\
VA1216 & 2020 & 1.67 & 4.94 & 0.945 & 9.17 \\
VA1217 & 2125 & 1.53 & 5.10 & 0.989 & 8.60 \\
VA1218 & 2015 & 1.51 & 5.16 & 1.025 & 9.60 \\
VA1220 & 2040 & 1.51 & 5.15 & 1.062 & 9.32 \\
VA1223 & 2025 & 1.65 & 4.51 & 1.023 & 9.66 \\
VA1224 & 1950 & 1.55 & 5.69 & 0.974 & 9.30 \\
VA1225 & 1800 & 1.62 & 5.50 & 1.093 & 9.55 \\
VA1226 & 2015 & 1.54 & 6.02 & 1.017 & 9.11 \\
VA1227 & 1860 & 1.51 & 6.85 & 1.028 & 9.36 \\
VA1228 & 1850 & 1.56 & 5.46 & 1.120 & 8.85 \\
VA1229 & 1865 & 1.50 & 5.01 & 1.101 & 8.97 \\
VA1230 & 2050 & 1.73 & 7.36 & 0.911 & 9.17 \\
VA1231 & 1980 & 1.66 & 5.11 & 0.917 & 9.30 \\
VA1232 & 2425 & 1.70 & 4.76 & 0.845 & 9.05 \\
VA1233 & 2170 & 1.51 & 6.61 & 0.935 & 9.67 \\
VA1235 & 2105 & 1.48 & 4.67 & 0.892 & 7.27 \\
VA1237 & 1910 & 1.61 & 5.53 & 1.037 & 9.47 \\
VA1238 & 1790 & 1.52 & 4.52 & 0.979 & 8.75 \\
VA1239 & 2050 & 1.57 & 4.72 & 1.022 & 9.38 \\
VA1240 & 1810 & 1.54 & 5.45 & 1.063 & 8.86 \\
VA1241 & 1980 & 1.67 & 5.26 & 0.926 & 9.19 \\
VA1242 & 2045 & 1.60 & 5.86 & 0.946 & 9.94 \\
VA1243 & 2045 & 1.61 & 6.30 & 1.012 & 9.35 \\
VA1244 & 2100 & 1.49 & 4.81 & 0.866 & 8.91 \\
VA1245 & 1945 & 1.64 & 5.23 & 1.063 & 8.80 \\
VA1246 & 1900 & 1.52 & 5.03 & 0.978 & 8.43 \\
VA1247 & 2075 & 1.52 & 4.85 & 0.934 & 8.05 \\
VA1248 & 1920 & 1.57 & 5.52 & 1.049 & 10.04 \\
VA1249 & 1860 & 1.49 & 5.58 & 0.951 & 8.68 \\
VA1251 & 2095 & 1.49 & 4.85 & 0.988 & 9.11 \\
VA1252 & 2095 & 1.53 & 6.12 & 0.908 & 9.77 \\
VA1253 & 1930 & 1.50 & 5.40 & 1.024 & 9.08 \\
VA1254 & 2005 & 2.00 & 6.65 & 0.986 & 10.06 \\
VA1255 & 2020 & 1.58 & 5.92 & 0.969 & 10.21 \\
VA1256 & 2075 & 1.59 & 5.65 & 0.852 & 9.05 \\
VA1257 & 2170 & 1.49 & 6.09 & 1.012 & 10.36 \\
VA1258 & 1910 & 1.55 & 5.04 & 0.968 & 10.02 \\
VA1259 & 1810 & 1.66 & 6.17 & 1.064 & 10.28 \\
VA1260 & 2230 & 1.61 & 6.17 & 0.953 & 10.51 \\
VA1263 & 1925 & 1.59 & 5.17 & 0.976 & 8.66 \\
VA1264 & 1880 & 1.54 & 6.33 & 1.010 & 10.21 \\
VA1265 & 2350 & 1.66 & 7.38 & 0.791 & 7.53 \\
VA1266 & 1980 & 1.58 & 5.46 & 1.050 & 9.60 \\
VA1267 & 1990 & 1.58 & 5.27 & 0.882 & 9.24 \\
VA1268 & 1850 & 1.49 & 5.08 & 0.972 & 8.55 \\
VA1269 & 2180 & 1.59 & 6.19 & 0.949 & 10.29 \\
VA1270 & 1970 & 1.54 & 4.76 & 0.895 & 9.30 \\
VA1271 & 1820 & 1.51 & 4.96 & 1.006 & 8.79 \\
VA1272 & 2000 & 1.59 & 5.95 & 0.962 & 10.61 \\
VA1273 & 2110 & 1.61 & 6.32 & 0.902 & 11.64 \\
VA1274 & 1955 & 1.59 & 4.99 & 0.946 & 9.33 \\
VA1276 & 2230 & 1.56 & 4.62 & 0.873 & 8.96 \\
VA1277 & 2095 & 1.53 & 5.15 & 0.867 & 9.36 \\
VA1278 & 2155 & 1.59 & 5.46 & 0.890 & 9.40 \\
VA1279 & 2085 & 1.65 & 5.83 & 0.934 & 9.01 \\
VA1280 & 1810 & 1.62 & 6.05 & 1.192 & 8.66 \\
VA1281 & 1905 & 1.58 & 5.54 & 1.082 & 8.72 \\
VA1283 & 2040 & 1.52 & 4.74 & 0.919 & 9.54 \\
VA1284 & 1880 & 1.59 & 5.72 & 0.970 & 8.56 \\
VA1285 & 2190 & 1.55 & 5.38 & 0.936 & 9.62 \\
VA1286 & 1890 & 1.61 & 4.77 & 0.991 & 9.30 \\
VA1287 & 1975 & 1.56 & 4.36 & 0.928 & 9.17 \\
VA1288 & 2065 & 1.58 & 5.65 & 0.878 & 9.23 \\
VA1289 & 2030 & 1.49 & 4.88 & 0.911 & 9.55 \\
VA1290 & 2070 & 1.53 & 5.62 & 0.950 & 8.94 \\
VA1291 & 1865 & 1.57 & 5.67 & 1.030 & 8.80 \\
VA1293 & 1845 & 1.58 & 5.37 & 0.952 & 8.22 \\
VA1295 & 1880 & 1.60 & 5.59 & 0.999 & 9.26 \\
VA1298 & 1855 & 1.64 & 5.64 & 1.000 & 9.90 \\
VA1299 & 2035 & 1.60 & 5.30 & 0.939 & 9.96 \\
VA1300 & 1900 & 1.57 & 5.00 & 1.026 & 8.92 \\
VA1301 & 1790 & 1.62 & 5.68 & 1.126 & 8.63 \\
VA1302 & 1975 & 1.58 & 5.67 & 1.008 & 8.72 \\
VA1303 & 1805 & 1.57 & 5.92 & 1.074 & 8.37 \\
VA1305 & 1855 & 1.64 & 4.83 & 1.118 & 8.48 \\
VA1307 & 1745 & 1.54 & 5.33 & 1.171 & 8.64 \\
VA1309 & 1840 & 1.65 & 5.28 & 1.083 & 9.52 \\
VA1312 & 1905 & 1.60 & 5.30 & 1.117 & 8.23 \\
VA1313 & 1805 & 1.58 & 5.16 & 1.134 & 8.25 \\
VA1314 & 1770 & 1.56 & 5.23 & 1.145 & 8.07 \\ \hline 
    \caption{All of the SPE characterization results at a gain of \gain. The TTS is the FWHM.}
    \label{tab:all_spe_results}
\end{longtable}

\clearpage 

\printbibliography 

@article{Anderson:2022lbb,
    author = "Anderson, T. and others",
    title = "{Eos: conceptual design for a demonstrator of hybrid optical detector technology}",
    eprint = "2211.11969",
    archivePrefix = "arXiv",
    primaryClass = "physics.ins-det",
    doi = "10.1088/1748-0221/18/02/P02009",
    journal = "JINST",
    volume = "18",
    number = "02",
    pages = "P02009",
    year = "2023"
}

@article{Theia:2019non,
    author = "Askins, M. and others",
    collaboration = "Theia",
    title = "{THEIA: an advanced optical neutrino detector}",
    eprint = "1911.03501",
    archivePrefix = "arXiv",
    primaryClass = "physics.ins-det",
    doi = "10.1140/epjc/s10052-020-7977-8",
    journal = "Eur. Phys. J. C",
    volume = "80",
    number = "5",
    pages = "416",
    year = "2020"
}

@article{Kaptanoglu:2019gtg,
    author = "Kaptanoglu, Tanner and Luo, Meng and Land, Ben and Bacon, Amanda and Klein, Josh",
    title = "{Spectral Photon Sorting For Large-Scale Cherenkov and Scintillation Detectors}",
    eprint = "1912.10333",
    archivePrefix = "arXiv",
    primaryClass = "physics.ins-det",
    doi = "10.1103/PhysRevD.101.072002",
    journal = "Phys. Rev. D",
    volume = "101",
    number = "7",
    pages = "072002",
    year = "2020"
}

@article{YehWbLS,
author = {Yeh, Minfang and others},
year = {2011},
pages = {51-56},
title = {A new water-based liquid scintillator and potential applications},
volume = {660},
journal = {Nucl. Instrum. Meth. A},
doi = {10.1016/j.nima.2011.08.040}
}

@article{Biller:2020uoi,
    author = "Biller, Steven D. and Leming, Edward J. and Paton, Josephine L.",
    title = "{Slow fluors for effective separation of Cherenkov light in liquid scintillators}",
    eprint = "2001.10825",
    archivePrefix = "arXiv",
    primaryClass = "physics.ins-det",
    doi = "10.1016/j.nima.2020.164106",
    journal = "Nucl. Instrum. Meth. A",
    volume = "972",
    pages = "164106",
    year = "2020"
}

@article{Kaptanoglu:2017jxo,
    author = "Kaptanoglu, Tanner",
    title = "{Characterization of the Hamamatsu 8'' R5912-MOD Photomultiplier Tube}",
    eprint = "1710.03334",
    archivePrefix = "arXiv",
    primaryClass = "physics.ins-det",
    doi = "10.1016/j.nima.2018.01.086",
    journal = "Nucl. Instrum. Meth. A",
    volume = "889",
    pages = "69--77",
    year = "2018"
}

@article{Brack:2012ig,
    author = "Brack, J. and others",
    title = "{Characterization of the Hamamatsu R11780 12 inch Photomultiplier Tube}",
    eprint = "1210.2765",
    archivePrefix = "arXiv",
    primaryClass = "physics.ins-det",
    doi = "10.1016/j.nima.2013.02.022",
    journal = "Nucl. Instrum. Meth. A",
    volume = "712",
    pages = "162--173",
    year = "2013"
}

@misc{r14688,
    author="Hamamatsu",
    title="{Photomultiplier tube R14688-100: Hamamatsu Photonics}", howpublished="\url{https://www.hamamatsu.com/us/en/product/optical-sensors/pmt/pmt_tube-alone/head-on-type/R14688-100.html}",
    note="[Accessed June 15, 2023]",
    year="2023"
}

@misc{h11934_datasheet,
      author="Hamamatsu",
      title="{Datasheet for R11265U SERIES / H11934 SERIES}",
      year="2019",
      note="[Accessed June 16, 2022]"
}

@article{Akindele:2023ixz,
    author = "Akindele, O. A. and others",
    title = "{Acceptance tests of Hamamatsu R7081 photomultiplier tubes}",
    eprint = "2306.09926",
    archivePrefix = "arXiv",
    primaryClass = "physics.ins-det",
    year = "2023"
}

@article{Akashi-Ronquest:2014jga,
    author = "Akashi-Ronquest, M. and others",
    title = "{Improving Photoelectron Counting and Particle Identification in Scintillation Detectors with Bayesian Techniques}",
    eprint = "1408.1914",
    archivePrefix = "arXiv",
    primaryClass = "physics.ins-det",
    doi = "10.1016/j.astropartphys.2014.12.006",
    journal = "Astropart. Phys.",
    volume = "65",
    pages = "40--54",
    year = "2015"
}

@article{SNO:2002tuh,
    author = "Ahmad, Q. R. and others",
    collaboration = "SNO",
    title = "{Direct evidence for neutrino flavor transformation from neutral current interactions in the Sudbury Neutrino Observatory}",
    eprint = "nucl-ex/0204008",
    archivePrefix = "arXiv",
    doi = "10.1103/PhysRevLett.89.011301",
    journal = "Phys. Rev. Lett.",
    volume = "89",
    pages = "011301",
    year = "2002"
}

@article{Super-Kamiokande:1998kpq,
    author = "Fukuda, Y. and others",
    collaboration = "Super-Kamiokande",
    title = "{Evidence for oscillation of atmospheric neutrinos}",
    eprint = "hep-ex/9807003",
    archivePrefix = "arXiv",
    reportNumber = "BU-98-17, ICRR-REPORT-422-98-18, UCI-98-8, KEK-PREPRINT-98-95, LSU-HEPA-5-98, UMD-98-003, SBHEP-98-5, TKU-PAP-98-06, TIT-HPE-98-09",
    doi = "10.1103/PhysRevLett.81.1562",
    journal = "Phys. Rev. Lett.",
    volume = "81",
    pages = "1562--1567",
    year = "1998"
}

@article{Borexino:2008dzn,
    author = "Arpesella, C. and others",
    collaboration = "Borexino",
    title = "{Direct Measurement of the Be-7 Solar Neutrino Flux with 192 Days of Borexino Data}",
    eprint = "0805.3843",
    archivePrefix = "arXiv",
    primaryClass = "astro-ph",
    doi = "10.1103/PhysRevLett.101.091302",
    journal = "Phys. Rev. Lett.",
    volume = "101",
    pages = "091302",
    year = "2008"
}

@article{DEAP:2019yzn,
    author = "Ajaj, R. and others",
    collaboration = "DEAP",
    title = "{Search for dark matter with a 231-day exposure of liquid argon using DEAP-3600 at SNOLAB}",
    eprint = "1902.04048",
    archivePrefix = "arXiv",
    primaryClass = "astro-ph.CO",
    doi = "10.1103/PhysRevD.100.022004",
    journal = "Phys. Rev. D",
    volume = "100",
    number = "2",
    pages = "022004",
    year = "2019"
}

@article{DayaBay:2012fng,
    author = "An, F. P. and others",
    collaboration = "Daya Bay",
    title = "{Observation of electron-antineutrino disappearance at Daya Bay}",
    eprint = "1203.1669",
    archivePrefix = "arXiv",
    primaryClass = "hep-ex",
    doi = "10.1103/PhysRevLett.108.171803",
    journal = "Phys. Rev. Lett.",
    volume = "108",
    pages = "171803",
    year = "2012"
}

@article{Barros:2015pjt,
    author = "Barros, N. and Kaptanoglu, T. and Kimelman, B. and Klein, J. R. and Moore, E. and Nguyen, J. and Stavreva, K. and Svoboda, R.",
    title = "{Characterization of the ETEL D784UKFLB 11 in. photomultiplier tube}",
    eprint = "1512.06916",
    archivePrefix = "arXiv",
    primaryClass = "physics.ins-det",
    doi = "10.1016/j.nima.2017.01.067",
    journal = "Nucl. Instrum. Meth. A",
    volume = "852",
    pages = "15--19",
    year = "2017"
}

@article{SNO:2021xpa,
    author = "Albanese, V. and others",
    collaboration = "SNO+",
    title = "{The SNO+ experiment}",
    eprint = "2104.11687",
    archivePrefix = "arXiv",
    primaryClass = "physics.ins-det",
    doi = "10.1088/1748-0221/16/08/P08059",
    journal = "JINST",
    volume = "16",
    number = "08",
    pages = "P08059",
    year = "2021"
}

@article{BILLER1999364,
title = {Measurements of photomultiplier single photon counting efficiency for the Sudbury Neutrino Observatory},
journal = {Nucl. Instrum. Meth. A},
volume = {432},
number = {2},
pages = {364-373},
year = {1999},
issn = {0168-9002},
doi = {https://doi.org/10.1016/S0168-9002(99)00500-8},
url = {https://www.sciencedirect.com/science/article/pii/S0168900299005008},
author = {S.D Biller and N.A Jelley and M.D Thorman and N.P Fox and T.H Ward},
}

@article{Land:2020oiz,
    author = "Land, B. J. and Bagdasarian, Z. and Caravaca, J. and Smiley, M. and Yeh, M. and Orebi Gann, G. D.",
    title = "{MeV-scale performance of water-based and pure liquid scintillator detectors}",
    eprint = "2007.14999",
    archivePrefix = "arXiv",
    primaryClass = "physics.ins-det",
    doi = "10.1103/PhysRevD.103.052004",
    journal = "Phys. Rev. D",
    volume = "103",
    number = "5",
    pages = "052004",
    year = "2021"
}

@article{AKCHURIN2007121,
title = {A study on ion initiated photomultiplier afterpulses},
journal = {Nucl. Instrum. Meth. A},
volume = {574},
number = {1},
pages = {121-126},
year = {2007},
issn = {0168-9002},
doi = {https://doi.org/10.1016/j.nima.2007.01.093},
url = {https://www.sciencedirect.com/science/article/pii/S0168900207001532},
author = {Nural Akchurin and Heejong Kim},
}

@article{Ma_2011,
	doi = {10.1016/j.nima.2010.11.095},
	url = {https://doi.org/10.1016%2Fj.nima.2010.11.095},
	year = {2011},
	publisher = {Elsevier {BV}},
	volume = {629},
	number = {1},
	pages = {93--100}, 
	author = {K.J. Ma and others},
	title = {Time and amplitude of afterpulse measured with a large size photomultiplier tube},
	journal = {Nucl. Instrum. Meth. A}
}

@article{Lei:2015lua,
    author = "Lei, Xiang-Cui and others",
    title = "{Evaluation of new large area PMT with high quantum efficiency}",
    eprint = "1504.03174",
    archivePrefix = "arXiv",
    primaryClass = "physics.ins-det",
    doi = "10.1088/1674-1137/40/2/026002",
    journal = "Chin. Phys. C",
    volume = "40",
    number = "2",
    pages = "026002",
    year = "2016"
}

@article{Wen:2019sik,
    author = "Wen, Liang-Jian and He, Miao and Wang, Yi-Fang and Cao, Jun and Liu, Shu-Lin and Heng, Yue-Kun and Qin, Zhong-Hua",
    title = "{A quantitative approach to select PMTs for large detectors}",
    eprint = "1903.12595",
    archivePrefix = "arXiv",
    primaryClass = "physics.ins-det",
    doi = "10.1016/j.nima.2019.162766",
    journal = "Nucl. Instrum. Meth. A",
    volume = "947",
    pages = "162766",
    year = "2019"
}

@article{Akerib:2012da,
    author = "Akerib, D. S. and others",
    title = "{An Ultra-Low Background PMT for Liquid Xenon Detectors}",
    eprint = "1205.2272",
    archivePrefix = "arXiv",
    primaryClass = "physics.ins-det",
    doi = "10.1016/j.nima.2012.11.020",
    journal = "Nucl. Instrum. Meth. A",
    volume = "703",
    pages = "1--6",
    year = "2013"
}

@misc{handbook,
    author="Hamamatsu",
    title="{Photomultiplier Tubes, Basics and Application, Fourth Addition}", howpublished="\url{https://www.hamamatsu.com/content/dam/hamamatsu-photonics/sites/documents/99_SALES_LIBRARY/etd/PMT_handbook_v4E.pdf}",
    note="[Accessed Aug. 10, 2023]",
    year="2023"
}

@misc{private_comm,
  author = "Hamamatsu",
  date = "Aug. 24, 2023",
  howpublished = "private communication"
}

@misc{ratpac,  
      author={Morgan Askins and others}, 
      title={ratpac-two},
      year={2023},
      publisher={Github},
    howpublished={https://github.com/rat-pac/ratpac-two}}

@article{LZ:2020fty,
    author = "Akerib, D. S. and others",
    collaboration = "LZ",
    title = "{The LUX-ZEPLIN (LZ) radioactivity and cleanliness control programs}",
    eprint = "2006.02506",
    archivePrefix = "arXiv",
    primaryClass = "physics.ins-det",
    doi = "10.1140/epjc/s10052-020-8420-x",
    journal = "Eur. Phys. J. C",
    volume = "80",
    number = "11",
    pages = "1044",
    year = "2020",
    note = "[Erratum: Eur.Phys.J.C 82, 221 (2022)]"
}

@article{LZ:2022ysc,
    author = "Aalbers, J. and others",
    collaboration = "LZ",
    title = "{Background determination for the LUX-ZEPLIN dark matter experiment}",
    eprint = "2211.17120",
    archivePrefix = "arXiv",
    primaryClass = "hep-ex",
    doi = "10.1103/PhysRevD.108.012010",
    journal = "Phys. Rev. D",
    volume = "108",
    number = "1",
    pages = "012010",
    year = "2023"
}

@article{surf_counting,
    author = {Tiedt, Douglas and Mount, Brianna and Rodriguez, Ayla},
    title = "{Counting facilities at the Black Hills Underground Campus}",
    journal = {AIP Conference Proceedings},
    volume = {2908},
    number = {1},
    pages = {020003},
    year = {2023},
    month = {09},
    issn = {0094-243X},
    doi = {10.1063/5.0161194},
    url = {https://doi.org/10.1063/5.0161194},
    eprint = {https://pubs.aip.org/aip/acp/article-pdf/doi/10.1063/5.0161194/18110386/020003\_1\_5.0161194.pdf},
}

\end{document}